\RequirePackage{lineno}

\documentclass[aps,prc,twocolumn,groupedaddress,showpacs,amsmath,amssymb,floatfix,superscriptaddress]{revtex4}
\usepackage{multirow}
\usepackage{graphicx}
\usepackage{times}
\usepackage[usenames]{color}
\newcommand{\PerTonDay}{(ton day)$^{-1}$}
\newcommand{\PerDay}{day$^{-1}$}
\newcommand{\PerKgYr}{(kg yr)$^{-1}$}

\begin{document}

\bibliographystyle{h-physrev3}

\title{Limit on Neutrinoless $\beta\beta$ Decay of $^{136}$Xe from the First Phase of KamLAND-Zen and Comparison with the Positive Claim in $^{76}$Ge}

% All university affiliations addresses go here:
\newcommand{\tohoku}{\affiliation{Research Center for Neutrino
    Science, Tohoku University, Sendai 980-8578, Japan}}
\newcommand{\osaka}{\affiliation{Graduate School of 
    Science, Osaka University, Toyonaka, Osaka 560-0043, Japan}}
\newcommand{\lbl}{\affiliation{Physics Department, University of
    California, Berkeley, and \\ Lawrence Berkeley National Laboratory, 
Berkeley, California 94720, USA}}
\newcommand{\colostate}{\affiliation{Department of Physics, Colorado
    State University, Fort Collins, Colorado 80523, USA}}
\newcommand{\ut}{\affiliation{Department of Physics and
    Astronomy, University of Tennessee, Knoxville, Tennessee 37996, USA}}
\newcommand{\tunl}{\affiliation{Triangle Universities Nuclear
    Laboratory, Durham, North Carolina 27708, USA and \\
Physics Departments at Duke University, North Carolina Central University,
and the University of North Carolina at Chapel Hill}}
\newcommand{\ipmu}{\affiliation{Kavli Institute for the Physics and Mathematics of the Universe (WPI),
University of Tokyo, Kashiwa, 277-8583, Japan}}
\newcommand{\nikhef}{\affiliation{Nikhef and the University of Amsterdam, Science Park, Amsterdam, the Netherlands}}
\newcommand{\washington}{\affiliation{Center for Experimental Nuclear Physics and Astrophysics, University of Washington, Seattle, Washington 98195, USA}}

%
% Note: some authors have joint appointments with IPMU (http://www.ipmu.jp/members/)
%
% Tohoku
\author{A.~Gando}\tohoku
\author{Y.~Gando}\tohoku
\author{H.~Hanakago}\tohoku
\author{H.~Ikeda}\tohoku
\author{K.~Inoue}\tohoku\ipmu
\author{K.~Ishidoshiro}\tohoku
\author{R.~Kato}\tohoku
\author{M.~Koga}\tohoku\ipmu
\author{S.~Matsuda}\tohoku
\author{T.~Mitsui}\tohoku
\author{D.~Motoki}\tohoku
\author{T.~Nakada}\tohoku
\author{K.~Nakamura}\tohoku\ipmu
\author{A.~Obata}\tohoku
\author{A.~Oki}\tohoku
\author{Y.~Ono}\tohoku
\author{M.~Otani}\tohoku
\author{I.~Shimizu}\tohoku
\author{J.~Shirai}\tohoku
\author{A.~Suzuki}\tohoku
\author{Y.~Takemoto}\tohoku
\author{K.~Tamae}\tohoku
\author{K.~Ueshima}\tohoku
\author{H.~Watanabe}\tohoku
\author{B.D.~Xu}\tohoku
\author{S.~Yamada}\tohoku
\author{H.~Yoshida}\tohoku

% IPMU
\author{A.~Kozlov}\ipmu

% Osaka
\author{S.~Yoshida}\osaka

% LBL and UC Berkeley
\author{T.I.~Banks}\lbl
\author{S.J.~Freedman}\ipmu\lbl
\author{B.K.~Fujikawa}\ipmu\lbl
\author{K.~Han}\lbl
\author{T.~O'Donnell}\lbl

% Colorado State
\author{B.E.~Berger}\colostate

% UT
\author{Y.~Efremenko}\ipmu\ut

% TUNL
\author{H.J.~Karwowski}\tunl
\author{D.M.~Markoff}\tunl
\author{W.~Tornow}\tunl

% Washington
\author{J.A.~Detwiler}\washington
\author{S.~Enomoto}\ipmu\washington

% NIKHEF
\author{M.P.~Decowski}\ipmu\nikhef

\collaboration{The KamLAND-Zen Collaboration}\noaffiliation

\date{\today}

\begin{abstract}

We present results from the first phase of the \mbox{KamLAND-Zen} double-beta decay experiment, corresponding to an exposure of 89.5\,kg yr of $^{136}$Xe. We obtain a lower limit for the neutrinoless double-beta decay half-life of $T_{1/2}^{0\nu} > 1.9 \times 10^{25}$\,yr at 90\% C.L. The combined results from \mbox{KamLAND-Zen} and \mbox{EXO-200} give $T_{1/2}^{0\nu} > 3.4 \times 10^{25}$\,yr at 90\% C.L., which corresponds to a Majorana neutrino mass limit of $\left<m_{\beta\beta}\right> < (120-250)\,{\rm meV}$ based on a representative range of available matrix element calculations. Using those calculations, this result excludes the Majorana neutrino mass range expected from the neutrinoless double-beta decay detection claim in $^{76}$Ge, reported by a part of the Heidelberg-Moscow Collaboration, at more than 97.5\% C.L.

\end{abstract}


\maketitle

Double-beta ($\beta\beta$) decay is a rare nuclear process observable in even-even nuclei for which ordinary beta decay is energetically forbidden or highly suppressed by large spin differences. Standard $\beta\beta$ decay proceeds by a second-order weak interaction emitting two electron anti neutrinos and two electrons ($2\nu\beta\beta$). If, however, the neutrino is a massive Majorana particle, $\beta\beta$ decay might also occur without the emission of neutrinos ($0\nu\beta\beta$). Observation of such a process would demonstrate that lepton number is not conserved in nature. Moreover, if the process is mediated by the exchange of a light left-handed neutrino, its rate increases with the square of the effective Majorana neutrino mass $\left<m_{\beta\beta}\right> \equiv \left| \Sigma_{i} U_{ei}^{2}m_{\nu_{i}} \right|$, and hence its measurement would provide information on the absolute neutrino mass scale. To date there has been only one claimed observation of $0\nu\beta\beta$ decay, in $^{76}$Ge~\cite{Klapdor2006}.

At present there are several operating experiments performing $0\nu\beta\beta$ decay searches with design sensitivities sufficient to test the Majorana neutrino mass implied by the claim in \cite{Klapdor2006} within a few years of running: GERDA with $^{76}$Ge, CUORE-0 with $^{130}$Te, and \mbox{EXO-200} and \mbox{KamLAND-Zen} with $^{136}$Xe. Among those experiments, \mbox{KamLAND-Zen} released its first $0\nu\beta\beta$ half-life limit, $T_{1/2}^{0\nu} > 5.7 \times 10^{24}$\,yr at 90\% C.L., based on a 27.4\,kg yr exposure~\cite{Gando2012a}. Although the sensitivity of this result was impeded by the presence of an unexpected background peak just above the 2.458\,MeV $Q$ value of $^{136}$Xe $\beta\beta$ decay, the Majorana neutrino mass sensitivity was similar to that in Ref.~\cite{Klapdor2006}. \mbox{EXO-200} later improved on this limit by a factor of 2.8~\cite{Auger2012}, constraining the result in \cite{Klapdor2006} for a number of nuclear matrix element (NME) calculations. 

As shown below, we have found the problematic background peak in the \mbox{KamLAND-Zen} spectrum to most likely come from metastable $^{110m}$Ag. We embarked recently on a purification campaign to remove this isotope.  Doing so required extracting the Xe from the detector, thus marking the end of the first phase of \mbox{KamLAND-Zen}. In this Letter we report on the full data set from the first phase of \mbox{KamLAND-Zen}, corresponding to an exposure of 89.5\,kg yr of $^{136}$Xe. This represents a factor of 3.2 increase over KamLAND-Zen's first result~\cite{Gando2012a}, and is also the largest exposure for a $\beta\beta$ decay isotope to date.
 
The \mbox{KamLAND-Zen} (KamLAND Zero-Neutrino Double-Beta Decay) experiment consists of \mbox{13\,tons} of Xe-loaded liquid scintillator~(Xe-LS) contained in a 3.08-m-diameter transparent nylon-based inner balloon (IB), suspended at the center of the KamLAND detector by film straps.  The IB is surrounded by 1\,kton  of liquid scintillator (LS) contained in a 13-m-diameter outer balloon. To detect scintillation light, 1,325 17-inch and 554 20-inch photomultiplier tubes (PMTs) are mounted on the stainless-steel containment tank (SST), providing 34\% photocathode coverage. The SST is surrounded by a \mbox{3.2-kton} water-Cherenkov detector for cosmic-ray muon identification. Details of the \mbox{KamLAND-Zen} detector are given in Ref.~\cite{Gando2012a}.

We report on data collected between October 12, 2011, and June 14, 2012. 
To address the possibility that impurities such as $^{110m}$Ag may be bound to suspended dust or fine particulate in the Xe-LS, in February 2012 we passed 37\,${\rm m}^{3}$ of the Xe-LS (corresponding to 2.3 full volume exchanges) through a 50\,nm PTFE-based filter. To facilitate analysis, we divided the data into two sets: one taken before (\mbox{DS-1}) and the other after (\mbox{DS-2}) the filtration. \mbox{DS-1} corresponds to the data set reported in Ref.~\cite{Gando2012b} except with the fiducial radius increased to 1.35\,m to optimize the $0\nu\beta\beta$ search, yielding a fiducial Xe-LS mass of 8.04\,tons. For \mbox{DS-2}, additional fiducial volume cuts were made around the siphoning hardware left in place after the filtration ended--namely, a 0.2-m-radius cylindrical cut along the length of the Teflon piping, as well as a 1.2-m-radius spherical cut around the stainless steel inlet at its tip. The total live time after removing periods of high background rate due to $^{222}$Rn daughters  introduced by the filtration is 213.4\,days. The live time, fiducial Xe-LS mass, Xe concentration, $^{136}$Xe mass, and exposure for the data sets are summarized in Table~\ref{table:fiducial}.

\begin{table}[t]
\begin{center}
\caption{\label{table:fiducial}Two data sets used in this $^{136}$Xe $0\nu\beta\beta$ decay analysis.}
\begin{tabular}{@{}*{4}{lccc}}
\hline
\hline
\hspace{3.0cm} & ~~~~~\mbox{DS-1}~~~~~ & ~~~~~\mbox{DS-2}~~~~~ & ~~~~Total~~~~ \\
\hline
live time (days) & 112.3 & 101.1 & 213.4 \\
fiducial Xe-LS mass (ton) & 8.04 & 5.55 & - \\
Xe concentration (wt\%) & 2.44 & 2.48 & - \\
$^{136}$Xe mass (kg) & 179 & 125 & - \\
$^{136}$Xe exposure (kg yr) & 54.9 & 34.6 & 89.5 \\
\hline
\hline
\end{tabular}
\vspace{-0.5cm}
\end{center}
\end{table}

Event vertex and energy are reconstructed based on the timing and charge distributions of photoelectrons recorded by the PMTs. Energy calibration is performed using $^{208}$Tl $\gamma$'s from a \mbox{${\rm ThO_{2}W}$} source~\cite{Gando2012a}, tagged $^{214}$Bi $\beta$'s and $\gamma$'s from $^{222}$Rn (\mbox{$\tau = 5.5$\,day}) introduced during the initial filling of the IB with Xe-LS, and 2.225\,MeV $\gamma$'s from capture of spallation neutrons by protons. Uncertainties from the nonlinear energy response due to scintillator quenching and Cherenkov light production are constrained by the calibrations. The energy scale variation was confirmed by the neutron-capture $\gamma$ data to be less than 1.0\% over the Xe-LS volume, and stable to within 1.0\% during the data taking period. The vertex resolution is $\sigma\sim 15\,{\rm cm}/\sqrt{E({\rm MeV})}$, and the energy resolution is $\sigma = (6.6 \pm 0.3)\%/\sqrt{E({\rm MeV})}$.

Double-beta decay events are selected by performing the following series of cuts: (i) The reconstructed vertex must be within the fiducial volume (FV) defined for each data set. (ii) Muons and events within 2\,ms after muons are rejected, imposing a dead time of $\sim$0.06\%. (iii) Events occurring within 3\,ms of each other are eliminated to avoid background from $^{214}$Bi-$^{214}$Po ($\tau$=237\,$\mu$s) decays. The dead time introduced by this coincidence cut is less than 0.1\%. (iv) Reactor $\overline{\nu}_{e}$'s identified by a delayed coincidence of positrons and neutron-capture $\gamma$'s as in \cite{Gando2011} are eliminated. (v) Poorly reconstructed events are rejected. These events are tagged using a vertex-time-charge (VTQ) discriminator which measures how well the observed PMT time-charge distributions agree with those expected based on the reconstructed vertex~\cite{Abe2011}. The VTQ cut reduces the selection efficiency by less than 0.1\%. The event selection criteria (ii-v) are the same as those described in detail in Ref.~\cite{Gando2012a}.  The cut (i) introduces an inefficiency from the balance of events reconstructing on either side of the fiducial boundary due to the vertex resolution. This inefficiency was estimated with a Monte Carlo simulation. The total efficiency for identifying $\beta\beta$ events above the analysis visible energy threshold ($E > 0.5$ MeV), is 99.8\% and 97.9\% for $0\nu\beta\beta$ and $2\nu\beta\beta$ decays, respectively. The uncertainty on the boundary effect correction is included in the systematic error on the FV cuts.

Nominally, the 1.35-m-radius FV for \mbox{DS-1} corresponds to $0.624 \pm 0.006$ of the total Xe-LS volume ($16.51 \pm 0.17\,{\rm m}^{3}$), or 179\,kg of $^{136}$Xe. The FV fraction is also estimated from the ratio of $^{214}$Bi events which pass the FV cuts to the total number in the entire Xe-LS volume after subtraction of the IB surface contribution, resulting is $0.620 \pm 0.007({\rm stat}) \pm 0.021({\rm syst})$. The difference in these values is taken as a measure of the systemic error on the vertex-defined FV. Combining the errors, we obtain a 3.9\% systematic uncertainty on the FV for \mbox{DS-1}. Similarly, the error for  \mbox{DS-2} is estimated to be 4.1\%. The total systematic uncertainties on the $\beta\beta$ decay half-life measurements for DS-1/DS-2 are 3.9\%/4.1\%~\cite{Gando2012b}, resulting from the quadrature sum of the uncertainties in the fiducial volume (3.9\%/4.1\%), enrichment level of $^{136}$Xe (0.05\%)~\cite{Gando2012a}, Xe concentration (0.34\%/0.37\%), detector energy scale (0.3\%)~\cite{Gando2012a}, and detection efficiency (0.2\%).

The main contributors to the $\beta\beta$ decay background can be divided into three categories: those from radioactive impurities in the Xe-LS; those from muon-induced spallation products; and those external to the Xe-LS, mainly from the IB material. The U and Th contaminations in the Xe-LS can be investigated by the delayed coincidence detection of $^{214}$Bi-$^{214}$Po and $^{212}$Bi-$^{212}$Po. Assuming secular equilibrium, the $^{238}$U and $^{232}$Th concentrations are estimated to be $(1.3 \pm 0.2) \times 10^{-16}$\,g/g and $(1.8 \pm 0.1) \times 10^{-15}$\,g/g, respectively. The $^{238}$U level reported in Ref.~\cite{Gando2012a} was overestimated due to slight contamination of $^{222}$Rn in early data, which can be removed. To allow for the possibility of decay chain nonequilibrium, however, the Bi-Po measurements are used to constrain only the rates for the $^{222}$Rn-$^{210}$Pb subchain of the $^{238}$U series and the $^{228}$Th-$^{208}$Pb subchain of the $^{232}$Th series, while other background rates in both series as well as a contribution from $^{85}$Kr are left unconstrained. 

Spallation neutrons are captured mainly on protons (2.225\,MeV) and $^{12}$C (4.946\,MeV) in organic scintillator components, and only rarely on $^{136}$Xe (4.026\,MeV) and $^{134}$Xe (6.364\,MeV), with fractions of the total captures, $9.5 \times 10^{-4}$ and $9.4 \times 10^{-5}$, respectively, for the latter two. The neutron capture product $^{137}$Xe ($\beta^{-}$, \mbox{$\tau=5.5$\,min}, \mbox{$Q=4.17$\,MeV}) is a potential background, but its expected rate is negligible in the current $0\nu\beta\beta$ search. For carbon spallation products, we expect event rates of $1.11 \pm 0.28$\,\PerTonDay\, and $(2.11 \pm 0.44) \times 10^{-2}$\,\PerTonDay\, from $^{11}$C ($\beta^{+}$, \mbox{$\tau=29.4$\,min}, \mbox{$Q=1.98$\,MeV}) and $^{10}$C ($\beta^{+}$, \mbox{$\tau = 27.8$\,s}, \mbox{$Q = 3.65$\,MeV}), respectively. There are no past experimental data for muon spallation of Xe,   but background from short-lived products of Xe with lifetimes of less than 100\,s is constrained from the study of muon time-correlated events~\cite{Gando2012a}.

By looking at events near the IB radius, we found that the IB, which was fabricated 100\,km from the Fukushima-I reactor, was contaminated by fallout from the Fukushima nuclear accident in March 2011~\cite{Gando2012a}. The dominant activities from this fallout are $^{134}$Cs ($\beta + \gamma$'s)  and $^{137}$Cs (0.662\,MeV $\gamma$), but they do not generate background in the energy region $2.2<E<3.0$\,MeV relevant to the $^{136}$Xe $0\nu\beta\beta$ decay search (i.e., the $0\nu\beta\beta$ window). In this region, the dominant IB contaminant is $^{214}$Bi ($\beta + \gamma$'s) from the U decay chain. The Cs and U are not distributed uniformly on the IB film. Rather, their activity appears to increase proportionally with the area of the film welding lines. This indicates that the dominant IB backgrounds may have been introduced during the welding process from dust containing both natural U and Fukushima fallout contaminants. The activity of the $^{214}$Bi on the IB drives the spherical fiducial radius in the analysis.

\begin{figure}[t]
\begin{center}
\vspace{-0.5cm}
\includegraphics[angle=270,width=1.0\columnwidth]{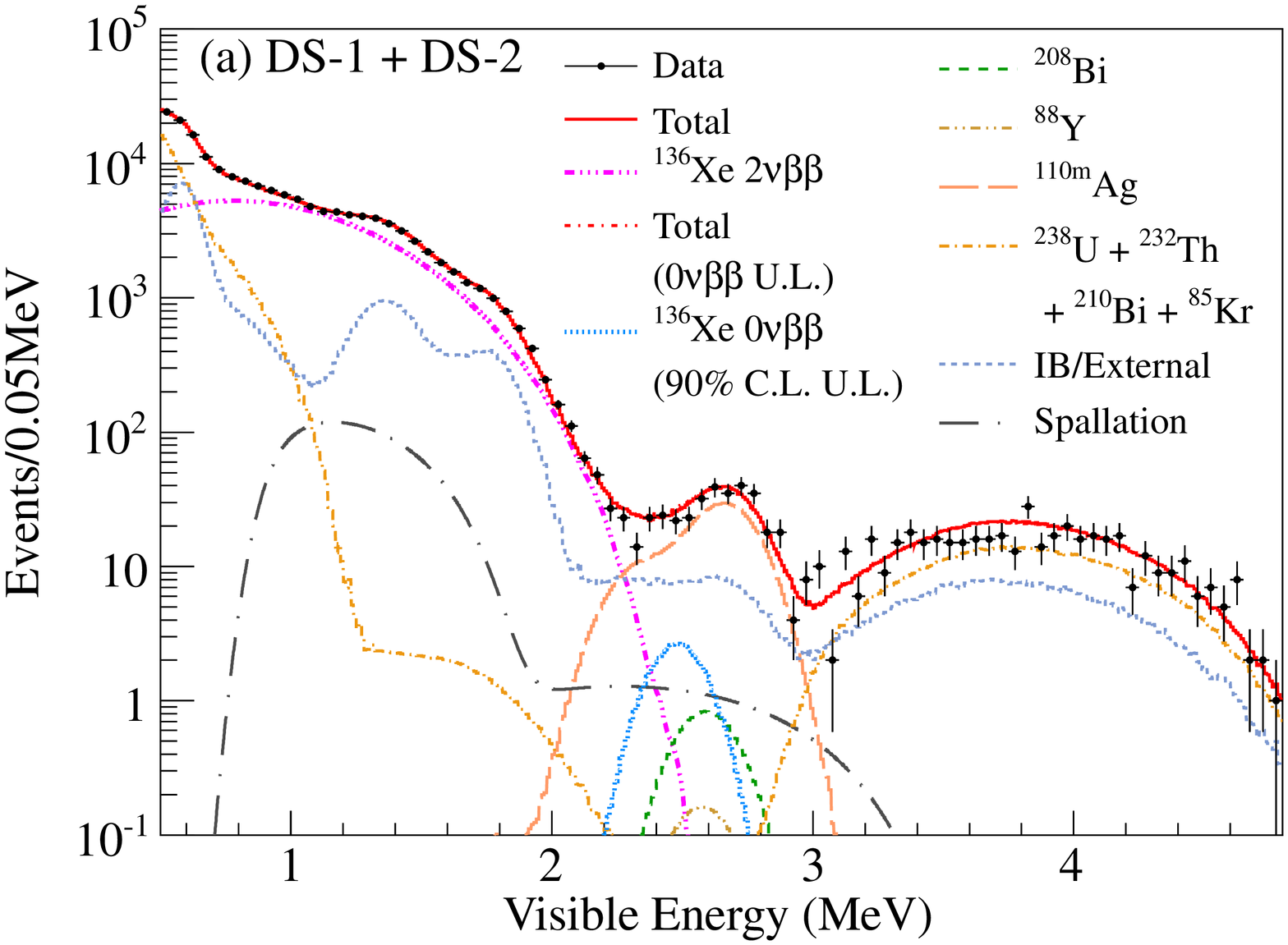}
\end{center}
\vspace{-0.5cm}
\begin{center}
\includegraphics[angle=270,width=1.0\columnwidth]{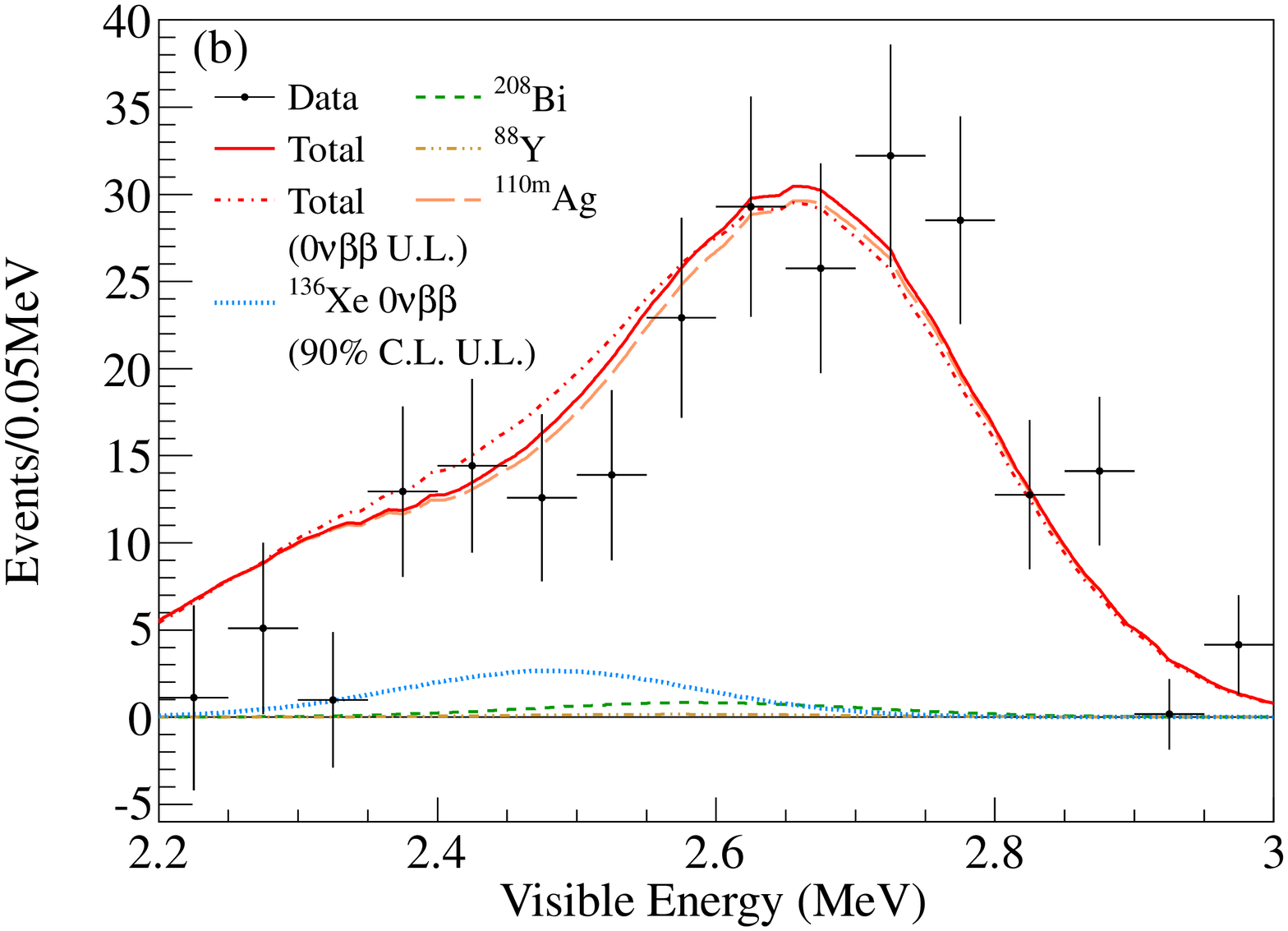}
\vspace{-0.5cm}
\end{center}
\caption[]{(a) Energy spectrum of selected candidate events together with the best-fit backgrounds and $2\nu\beta\beta$ decays, and the 90\% C.L. upper limit for $0\nu\beta\beta$ decays, for the combined data from \mbox{DS-1} and \mbox{DS-2}; the fit range is $0.5 < E < 4.8\,{\rm MeV}$. (b) Closeup of (a) for $2.2 < E < 3.0\,{\rm MeV}$ after subtracting known background contributions.}
\label{figure:energy}
\end{figure}

In the combined \mbox{DS-1} and \mbox{DS-2} data set, a peak can also be observed in the IB backgrounds located in the $0\nu\beta\beta$ window on top of the $^{214}$Bi contribution, similar in energy to the peak found within the fiducial volume. To explore this activity we performed two-dimensional fits in $R$ and energy, assuming that the only contributions on the IB are from $^{214}$Bi and $^{110m}$Ag. Floating the rates from background sources uniformly distributed in the Xe-LS, the fit results for the $^{214}$Bi and $^{110m}$Ag event rates on the IB are $19.0 \pm 1.8$\,\PerDay and $3.3 \pm 0.4$\,\PerDay, respectively, for \mbox{DS-1}, and $15.2 \pm 2.3$\,\PerDay and $2.2 \pm 0.4$\,\PerDay for \mbox{DS-2}. The $^{214}$Bi rates are consistent between \mbox{DS-1} and \mbox{DS-2} given the different fiducial volume selection, while the $^{110m}$Ag rates are consistent with the decay time of this isotope. The rejection efficiencies of the FV cut $R < 1.35\,{\rm m}$ against $^{214}$Bi and $^{110m}$Ag on the IB are $(96.8 \pm 0.3)\%$ and $(93.8 \pm 0.7)\%$, respectively, where the uncertainties include the uncertainty in the IB position.

\begin{figure}[b]
\begin{center}
\vspace{0.0cm}
\includegraphics[angle=270,width=0.9\columnwidth]{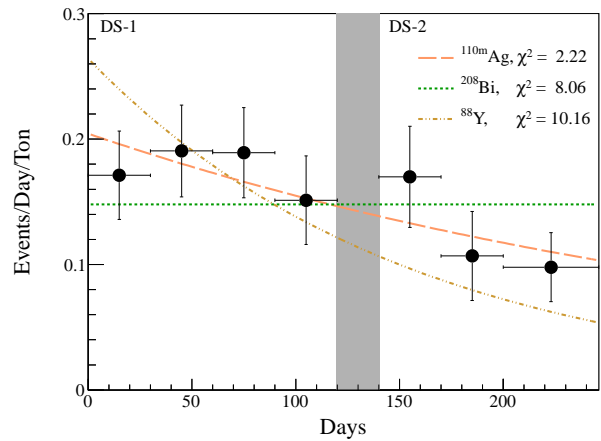}
\vspace{-0.4cm}
\end{center}
\caption[]{Event rate variation in the energy region $2.2 < E < 3.0\,{\rm MeV}$ ($^{136}$Xe $0\nu\beta\beta$ window) after subtracting known background contributions. The three fitted curves correspond to the hypotheses that all events in the $0\nu\beta\beta$ window are from $^{110m}$Ag (dashed line), $^{208}$Bi (dotted line), or $^{88}$Y (double-dot-dashed line). The gray band indicates the Xe-LS filtration period; no reduction in the fitted isotope is assumed for the $\chi^{2}$ calculation.}
\label{figure:time}
\end{figure}

The energy spectra of selected candidate events for \mbox{DS-1} and \mbox{DS-2} are shown in Fig.~\ref{figure:energy}. The $\beta\beta$ decay rates are estimated from a likelihood fit to the binned energy spectrum between 0.5 and 4.8\,MeV for each data set. The background rates described above are floated but constrained by their estimated values, as are the detector energy response model parameters. As discussed in Ref.~\cite{Gando2012a}, contributions from $^{110m}$Ag ($\beta^{-}$ decay, \mbox{$\tau=360$\,day}, \mbox{$Q = 3.01$\,MeV}), $^{88}$Y (EC decay, \mbox{$\tau=154$\,day}, \mbox{$Q = 3.62$\,MeV}), $^{208}$Bi (EC decay, \mbox{$\tau=5.31 \times 10^{5}$\,yr}, \mbox{$Q = 2.88$\,MeV}), and $^{60}$Co ($\beta^{-}$ decay, \mbox{$\tau = 7.61$\,yr}, \mbox{$Q = 2.82$\,MeV}) are considered as potential background sources in the $0\nu\beta\beta$ region of interest. The increased exposure time of this data set allows for improved constraints on the identity of the background due to the different lifetimes of the considered isotopes. Fig.~\ref{figure:time} shows the event rate time variation in the energy range $2.2 < E < 3.0\,{\rm MeV}$, which exhibits a strong preference for the lifetime of $^{110m}$Ag, if the filtration is assumed to have no effect. Allowing for the $^{110m}$Ag levels between \mbox{DS-1} and \mbox{DS-2} to float, the estimated removal efficiency of $^{110m}$Ag is $(1 \pm 19)\%$, indicating that the Xe-LS filtration was not effective in reducing the background. In the fit to extract the $0\nu\beta\beta$ limit we include all candidate sources in the Xe-LS, considering the possibility of composite contributions and allowing for independent background rates before and after the filtration.

The best-fit event rate of $^{136}$Xe $2\nu\beta\beta$ decays is $82.9 \pm 1.1 ({\rm stat}) \pm 3.4 ({\rm syst})$\,\PerTonDay for \mbox{DS-1}, and $80.2 \pm 1.8 ({\rm stat}) \pm 3.3 ({\rm syst})$\,\PerTonDay for \mbox{DS-2}. 82\% of the $2\nu\beta\beta$ spectrum falls within the analysis visible energy window ($0.5 < E < 4.8\,{\rm MeV}$). These results are consistent within the uncertainties, and both data sets indicate a uniform distribution of the Xe throughout the Xe-LS. They are also consistent with \mbox{EXO-200}~\cite{Auger2012} and that obtained with a smaller exposure~\cite{Gando2012b}, which requires the FV cut $R < 1.2\,{\rm m}$ to avoid the large $^{134}$Cs backgrounds on the IB, more appropriate for the $2\nu\beta\beta$ analysis.

The best-fit $^{110m}$Ag rates in the Xe-LS are $0.19 \pm 0.02$\,\PerTonDay and $0.14 \pm 0.03$\,\PerTonDay for \mbox{DS-1} and \mbox{DS-2}, respectively, indicating a dominant contribution of $^{110m}$Ag in the $0\nu\beta\beta$ region. The next largest background is $^{214}$Bi on the IB remaining after the FV cut, while $^{208}$Bi, $^{88}$Y, and $^{60}$Co have at most minor contributions. The 90\% C.L. upper limits on the number of $^{136}$Xe $0\nu\beta\beta$ decays are $<$16\,events and $<$8.7\,events for \mbox{DS-1} and \mbox{DS-2}, respectively. Combining the results, we obtain a 90\% C.L. upper limit of $<$0.16\,\PerKgYr in units of $^{136}$Xe exposure, or $T_{1/2}^{0\nu} > 1.9 \times 10^{25}$\,yr (90\% C.L.). This corresponds to a factor of 3.3 improvement over the first \mbox{KamLAND-Zen} result~\cite{Gando2012a}. The hypothesis that backgrounds from $^{88}$Y, $^{208}$Bi, and $^{60}$Co are absent marginally increases the limit to $T_{1/2}^{0\nu} > 2.0 \times 10^{25}$\,yr (90\% C.L.). A Monte Carlo simulation of an ensemble of experiments based on the best-fit background spectrum indicates a sensitivity~\cite{Feldman1998} of $1.0 \times 10^{25}$\,yr. The chance of obtaining a limit equal to or stronger than that reported here is 12\%.

\begin{figure}[t]
\begin{center}
\vspace{0.0cm}
\includegraphics[angle=270,width=1.0\columnwidth]{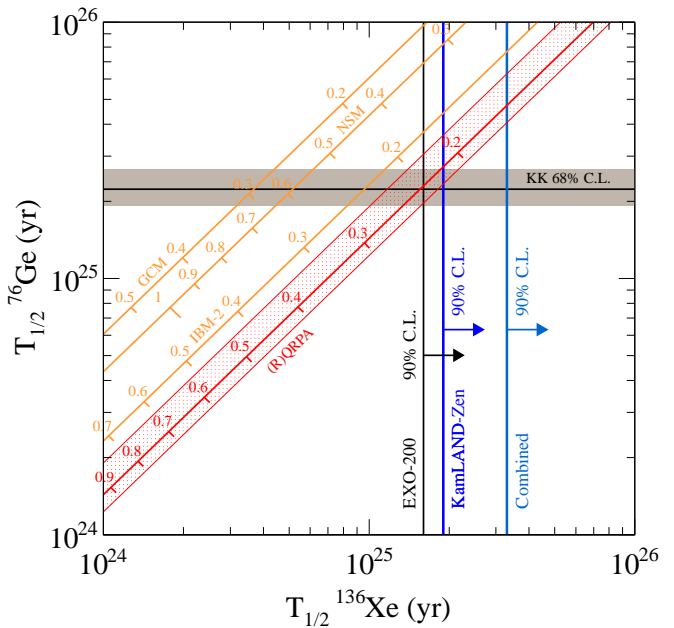}
\vspace{-0.5cm}
\end{center}
\caption[]{Experimental results on $0\nu\beta\beta$ decay half-life ($T_{1/2}^{0\nu}$) in $^{76}$Ge and $^{136}$Xe. The 68\% C.L. limit from the claim in Ref. ~\cite{Klapdor2006} is indicated by the gray band. The limits for \mbox{KamLAND-Zen} (this work), \mbox{EXO-200}~\cite{Auger2012}, and their combination are shown at 90\% C.L. The correlation between the $^{76}$Ge and $^{136}$Xe half-lives predicted by various NME calculations~\cite{Rodriguez2010,Menendez2009,Barea2012,Faessler2012} is drawn as diagonal lines together with the $\left<m_{\beta\beta}\right>$\,(eV) scale. The band for QRPA and RQRPA represents the range of these NME under the variation of model parameters.}
\label{figure:halflife}
\end{figure}

A combination of the limits from \mbox{KamLAND-Zen} and \mbox{EXO-200}, constructed by a $\chi^{2}$ test tuned to reproduce the result in Ref.~\cite{Auger2012}, gives $T_{1/2}^{0\nu} > 3.4 \times 10^{25}$\,yr (90\% C.L.). The combined measurement has a sensitivity of $1.6 \times 10^{25}$\,yr, and the probability of obtaining a stronger limit is 7\%. From the combined half-life limit, we obtain a 90\% C.L. upper limit of $\left<m_{\beta\beta}\right> < (120-250)\,{\rm meV}$ considering various NME calculations~\cite{Rodriguez2010,Menendez2009,Barea2012,Faessler2012}. The constraint from this combined result on the detection claim in Ref.~\cite{Klapdor2006} is shown in Fig.~\ref{figure:halflife} for different NME estimates. We find that the combined result for $^{136}$Xe refutes the $0\nu\beta\beta$ detection claim in $^{76}$Ge at $>$97.5\% C.L. for all NME considered assuming that $0\nu\beta\beta$ decay proceeds via light Majorana neutrino exchange. While the statistical treatment of the NME uncertainties is not straightforward, even if we apply the uncertainties and correlations in Ref.~\cite{Faessler2009}, which assumes a statistical distribution of the NME for various (R)QRPA models and does not include a tuning of the parameter $g_{pp}$ for $^{136}$Xe based on its measured $2\nu\beta\beta$ half-life, we find the rejection significance is still 95.6\% C.L.

The \mbox{KamLAND-Zen} result is still limited by the background from $^{110m}$Ag. The two leading hypotheses to explain its presence in the Xe-LS are (i) IB contamination during fabrication by Fukushima-I fallout and (ii) cosmogenic production by Xe spallation~\cite{Gando2012a}. While the distribution of Cs isotopes is consistent with IB contamination during fabrication, hypothesis of the adsorption of cosmogenically produced $^{110m}$Ag onto the IB still cannot be rejected. Improved statistics on the distribution of $^{110m}$Ag on the IB may help reveal the source of the contamination. In the meantime, we have removed the Xe from the Xe-LS by vacuum extraction and verified that the $^{110m}$Ag rate in the LS remains at its present level. We are proceeding to distill the LS to remove the $^{110m}$Ag, while we also pursue options for IB replacement and further detector upgrades.

In summary, we have performed the most stringent test to date on the claimed observation of  $0\nu\beta\beta$ decay in $^{76}$Ge~\cite{Klapdor2006}. Combining the limits on $^{136}$Xe $0\nu\beta\beta$ decay by \mbox{KamLAND-Zen} and \mbox{EXO-200}, we find that the Majorana mass range expected from the claimed $^{76}$Ge $0\nu\beta\beta$ decay half-life is excluded at $>$97.5\% C.L. for a representative range of nuclear matrix element estimations. \mbox{KamLAND-Zen} and \mbox{EXO-200} demonstrate that we have arrived at an exciting new era in the field, and that the technology needed to judge the claimed $^{76}$Ge $0\nu\beta\beta$ decay with other nuclei has been achieved.

The \mbox{KamLAND-Zen} experiment is supported by the Grant-in-Aid for Specially Promoted Research under grant 21000001 of the Japanese Ministry of Education, Culture, Sports, Science and Technology; the World Premier International Research Center Initiative (WPI Initiative), MEXT, Japan; Stichting FOM in the Netherlands; and under the US Department of Energy (DOE) Grant No. DE-AC02-05CH11231, as well as other DOE grants to individual institutions. The Kamioka Mining and Smelting Company has provided service for activities in the mine.

This work is dedicated to the memory of Stuart Freedman, a great scientist, and a dear colleague and friend.

\bibliography{}

\end{document}